\documentclass[a4paper,11pt]{article}
\usepackage[numbers,sort&compress]{natbib}
\usepackage{pos}

\usepackage{amsmath}
\usepackage{epsfig}
\usepackage{graphicx}
\usepackage{amsfonts}
\usepackage{mathrsfs}
\usepackage{pifont}
\usepackage{relsize}
\usepackage{mathtools}
\def\slashchar#1{\setbox0=\hbox{$#1$}     		% set a box for #1
   \dimen0=\wd0                                 	% and get its size
   \setbox1=\hbox{/} \dimen1=\wd1               	% get size of /
   \ifdim\dimen0>\dimen1                        	% #1 is bigger
      \rlap{\hbox to \dimen0{\hfil/\hfil}}      	% so center / in box
      #1                                        	% and print #1
   \else                                        	% / is bigger
      \rlap{\hbox to \dimen1{\hfil$#1$\hfil}}   	% so center #1
      /                                         	% and print /
   \fi}

\renewcommand{\vec}{\boldsymbol}
\newcommand{\beq}{\begin{equation}}
\newcommand{\eeq}{\end{equation}}
\newcommand{\bea}{\begin{eqnarray}}
\newcommand{\eea}{\end{eqnarray}}
\newcommand{\baa}{\begin{array}}
\newcommand{\eaa}{\end{array}}

\def\eq#1{{Eq.~(\ref{#1})}}

\def\fig#1{{Fig.~\ref{#1}}}
\newcommand{\De}{\Delta}

\newcommand{\bas}{\bar{\alpha}_S}
\newcommand{\Lb}{\left(}
\newcommand{\Rb}{\right)}
\numberwithin{equation}{section}

\title{Confronting impact-parameter dependent model in next-to-leading order of perturbative QCD with combined HERA data}
\author*[a]{Jos\'e Garrido}
\author[b,c]{Michael Roa}
\author[b]{Miguel Guevara}

\affiliation[a]{Departamento de F\'isica, Universidad T\'ecnica Federico Santa Mar\'ia,  Avda. Espa\~na 1680, Casilla 110-V, Valpara\'iso, Chile}

\affiliation[b]{Facultad de Ingenier\'ia, Laboratorio DataScience, Universidad de Playa Ancha, Leopoldo Carvallo 270, Valpara\'iso, Chile}

\affiliation[c]{Centro de Estudios Avanzados, Universidad de Playa Ancha, Traslavi\~{n}a 450, Vi\~{n}a del Mar, Chile}

\emailAdd{jose.garridom@sansano.usm.cl}
\emailAdd{michael.roa@upla.cl}
\emailAdd{miguel.guevara@upla.cl}

\abstract{In this talk, we present the CGC/saturation approach of Ref.~\cite{CLMS} and its parameters determined from the combined HERA data. This model features an analytical solution for the non-linear Balitsky-Kovchegov (BK) evolution equation and the exponential behavior of the saturation momentum on the impact parameter $b$-dependence, characterized by $Q_s\propto \exp(-mb)$. We compare our results with experimental data at small-$x$, including the proton structure function $F_2$, charm structure function $F_2^{c\bar{c}}$, and exclusive vector meson production. The model shows good agreement across a wide kinematic range. Our findings support using this approach for reliable predictions in upcoming experiments like the Electron-Ion Collider (EIC) and the LHeC.}

\FullConference{31st International Workshop on Deep Inelastic Scattering (DIS2024)\\
 8–12 April 2024\\
Grenoble, France\\}

\begin{document}
\maketitle

\section{Introduction}

%%%%%%%%%%%%%%%%%%%%%%%%%%%%%%%%%%%%%%%%%%%%%%%%%%%%
The Color Glass Condensate (CGC)/saturation effective field theory~\cite{KOLEB} provides an effective description of high-energy QCD processes. Despite its success, leading-order CGC predictions show high-energy growth inconsistent with experimental data. To resolve this, Ref.~\cite{CLMS} introduced corrections to the BK equation~\cite{BK}, improving predictions for high-energy scattering. The model accurately reproduces the scattering amplitude's large impact parameter $b$-dependence, ensuring compliance with the Froissart theorem~\cite{FROI}, which is crucial for realistic high-energy predictions. This approach aligns with geometric scaling and perturbative QCD for large momentum transfers. We fit the CGC/saturation dipole model to precise HERA DIS data~\cite{HERA1,HERA2}, extracting phenomenological parameters and comparing theoretical predictions to experimental results for various observables such as the proton structure function $F_2$ and exclusive vector meson production. This study enhances the predictive power of the CGC/saturation framework, providing valuable insights for future experiments like the EIC and LHeC. 
%%%%%%%%%%%%%%%%%%%%%%%%%%%%%%%%%%%%%%%%%%%%%%%%%%%%%%%%%%%%%%%%%%%%%%%

%%%%%%%%%%%%%%%%%%%%%%%%%%%%%%%%%%%%%%%%%%%%%%%%%%%%%%%%%%%%%%%%%%%%%%%
\section{Inclusive and Exclusive Processes}

%%%%%%%%%%%%%%%%%%%%%%%%%%%%%%%%%%%%%%%%%%%%%%%%%%%%%%%%%%%%%%%%%%%%%%%

The observables in deep inelastic scattering (DIS) can be expressed through the following scattering amplitudes
\beq
N_{L,T}\Lb Q, Y; b\Rb = \int \frac{d^2 r}{4\pi} \int^1_0 d z \,|\Psi^{\gamma^*}_{L,T}\Lb Q, r, z\Rb|^2 \,N\Lb r, Y; b\Rb
\eeq
Here, $Y = \ln(1/x_{Bj})$, $x_{Bj}$ is the Bjorken $x$, $z$ is the fraction of the virtual photon momentum carried by the quark, and $Q$ is the photon virtuality. The $|\Psi^{\gamma^*}_{L,T}\Lb Q, r, z\Rb|^2$ are the overlap wave-functions of the photon, detailed in Ref.~\cite{KOLEB}.
The main DIS observables are expressed as:
\begin{subequations}
 \bea 
 \sigma_{T,L}^{\gamma^*p}(Q^2,x) &\,\, =\,\,& 2 \int d^2 b \,\,N_{T,L}\Lb Q,Y;b\Rb; \label{SIGMA}\\
 F_2\Lb Q^2, x\Rb &=& \frac{Q^2}{4\,\pi^2\,\alpha_{\rm e.m.}} \big[\sigma_T^{\gamma^*p}(Q^2,x)\,\,+\,\,\sigma_L^{\gamma^*p}(Q^2,x)\big];\label{F2}\\
  F^{c\bar{c}}_2\Lb Q^2, x\Rb &=& \frac{Q^2}{4\,\pi^2\,\alpha_{\rm e.m.}} \Big[\sigma^{c\bar{c},\gamma^*p}_T(Q^2,x)\,\,+\,\,\sigma^{c\bar{c},\gamma^*p}_L(Q^2,x)\Big];\label{FCC}\\
  F_L\Lb Q^2, x\Rb &=& \frac{Q^2}{4\,\pi^2\,\alpha_{\rm e.m.}} \,\sigma_L^{\gamma^*p}(Q^2,x);\label{L} 
 \eea
\end{subequations} 
The reduced cross-section $\sigma_r$ is expressed in terms of the inclusive proton structure functions $F_2$ and $F_L$ as
\beq \label{SIGMAR}
\sigma_r(Q^2,x,y)=F_2(Q^2,x)-\frac{y^2}{1+(1-y)^2}F_L(Q^2,x)
\eeq
For exclusive diffractive processes $\gamma^{*} + p \rightarrow E + p$ (where $E$ is a real photon in DVCS or a vector meson), the amplitude and differential cross-section are given by:

\beq
{\cal{A}}^{\gamma^{*}p \rightarrow E p}_{T,L}(x,Q,\De) = 2i \int d^{2}\vec{r} \int_{0}^{1} \frac{dz}{4\pi} \int d^{2}\vec{b} (\Psi_{E}^{*}\Psi)_{T,L} \mathrm{e}^{-i[ \vec{b}-(\frac{1}{2}-z)\vec{r}]\cdot\vec{\De}}\,N\Lb r, Y; b\Rb
\eeq

\beq
\frac{d\sigma_{T,L}^{\gamma^{*}p \rightarrow E p}}{dt} = \frac{1}{16\pi}\left|{\cal{A}}^{\gamma^{*}p \rightarrow E p}_{T,L}\right|^{2} (1+\beta^2)
 \eeq

where the expressions for the overlap wave functions can be found in Ref.~\cite{SATMOD3} and $\beta$ accounts for the ratio of the real to imaginary parts of the scattering amplitude. Two main observables used in the analysis are the total cross-section $\sigma^{\gamma^{*}p \rightarrow E p}_{T,L}$, defined as:

\beq\label{SIGMAEXC}
\sigma^{\gamma^{*}p \rightarrow E p}_{T,L} = \int dt \frac{d\sigma_{T,L}^{\gamma^{*}p \rightarrow E p}}{dt}
\eeq

and the slope parameter $B_D$, obtained from the logarithmic derivative of the differential cross-section at $t=0$:

\beq\label{B_D}
B_D = \lim_{t\rightarrow 0} \frac{d}{dt}\ln\left(\frac{d\sigma_{T,L}^{\gamma^{*}p \rightarrow E p}}{dt}\right).
\eeq
%%%%%%%%%%%%%%%%%%%%%%%%%%%%%%%%%%%%%%%%%%%%%%%%%%%%%%%%%%%%%%%%%%%
\section{CGC/saturation dipole model}
%%%%%%%%%%%%%%%%%%%%%%%%%%%%%%%%%%%%%%%%%%%%%%%%%%%%%%
\subsection{$q\bar{q}$ dipole-proton scattering amplitude}
%%%%%%%%%%%%%%%%%%%%%%%%%%%%%%%%%%%%%%%%%%%%%%%%%%%%%%
To compute the total cross-sections, proton structure functions in DIS, exclusive diffractive vector meson production, and DVCS, we use the $q\bar{q}$ dipole-proton forward scattering amplitude. Ref.~\cite{CLMS} provides an analytical solution for the nonlinear BK equation at NLO BFKL kernel in the saturation domain, incorporating non-linear evolution and resummation procedures~\cite{SALAM}. The color $q\bar{q}$ dipole-proton scattering amplitude is given by: 

\bea \label{NZ}
N\Lb z\Rb = \left\{\begin{array}{l} N_0\,e^{z\bar{\gamma}} \ \ \ \ \ \ \ \ \ \ \ \ \ \ \ ~~~~~ \,\mbox{for}\ \ \tau\leq1;\\\ \\
a\,\Big( 1 - e^{- \Omega\Lb z \Rb}\Big) + \Lb 1 - a\Rb\frac{\Omega\Lb z \Rb}{1 + \Omega\Lb z \Rb}~~~~\mbox{for} \ \ \tau > 1; \end{array}
\right.
\eea
where $a = 0.65$, $\Omega(z)$ is:

\beq \label{OMEGA}
\Omega(z) = \Omega_0 \Bigg\{ \cosh \left(\sqrt{\sigma } z\right) + \frac{\bar{\gamma}}{\sqrt{\sigma}} \sinh \left(\sqrt{\sigma} z\right)\Bigg\},\;\sigma=\frac{\bas}{\lambda(1+\bas)};
\eeq

$N_0$ is a parameter fitted from data. The variable $z$ is defined as:

\beq \label{xiformula}
z=\ln\,(r^2Q^2_s(Y,b)).
\eeq 
The critical anomalous dimension and energy behavior of the saturation scale are:

\beq \label{gammaold}
\bar{\gamma} = \bar{\gamma}_\eta(1+\lambda_\eta),~~~~~~\bar{\gamma}_\eta=\sqrt{2+\bas}-1;
\eeq 

\beq \label{lambda}
\lambda = \frac{\lambda_\eta}{1+\lambda_\eta},~~~~\lambda_\eta = \frac{1}{2}\,\frac{\bas}{3+\bas-2\sqrt{2+\bas}};
\eeq

where $\eta=Y-\xi$ and $Y=\ln(1/x)$, $\xi=\ln(r^2Q_s^2(Y=0,\mathbf{b}))$. Expanding the linear solution of \eq{NZ} to $\tau<1$, we replace $\bar{\gamma}$ by:

\beq \label{gammanew}
\bar{\gamma}\rightarrow\bar{\gamma}+\frac{\ln(1/\tau)}{2\kappa\lambda Y},~~~~\kappa=\frac{\chi''(\bar{\gamma})}{\chi'(\bar{\gamma})}=\frac{\frac{d^2\omega(\bar{\gamma}_\eta)}{d\bar{\gamma}_\eta^2}}{\frac{d\omega(\bar{\gamma}_\eta)}{d\bar{\gamma}_\eta}}
\eeq

this equation was derived in Ref.~\cite{IIML} and has demonstrated a strong agreement with the experimental data for $x\leq0.01$.%%%%%%%%%%%%%%%%%%%%%%%%%%%%%%%%%%%%%%%%%%%%%%%%%%%%%%%%%%%%%%%%%%%%%%%%%%%%%%%%
\subsection{Impact parameter dependence of the saturation scale}
%%%%%%%%%%%%%%%%%%%%%%%%%%%%%%%%%%%%%%%%%%%%%%%%%%%%%%%%%%%%%%%%%%%%%%%%%%%%%%%%

We introduce a phenomenological parameter, $N_0$, determining the scattering amplitude at $r^2 Q^2_s = 1$. Initial conditions at $Y=0$ specify the saturation scale at $b=0$ and its dependence on $b$. We use the following expression to represent the saturation momentum~\cite{CLP}:
\bea
Q^{2}_s(Y, b) = Q^{2}_0 (m\,b\, K_{1}(m\,b))^{1/\bar{\gamma}} e^{\lambda Y},
\eea
this ensures the correct large $b$ behavior. The parameters $Q^{2}_0$ and $m$ are fitted to data, with $Q^{2}_0 \in [0.15, 0.25]\,\rm{GeV}^2$ and $m \in [0.4, 0.85]\,\rm{GeV}$. The value of $\lambda=0.2$ is needed to describe DIS data, with $\bas \approx 0.1$ treated as a free parameter.

%%%%%%%%%%%%%%%%%%%%%%%%%%%%%%%%%%%%%%%%%%%%%%%%%%%%%%%%%%%%%%%%%%%
\section{Numerical results and discussion}
%%%%%%%%%%%%%%%%%%%%%%%%%%%%%%%%%%%%%%%%%%%%%%%%%%%%%%%%%%%%%%%%%%%

The model uses four parameters fitted to H1 and ZEUS data~\cite{HERA1}, and the fit range ensures BK equation validity, with $\chi^2$ minimization including systematic and statistical uncertainties. 

\begin{table}[ht]
{\footnotesize
\begin{tabular}{||l|l|l|l|l|l||}
\hline
\hline
\multicolumn{4}{||c|}{Dipole amplitude} & \multicolumn{1}{c||}{Minimization} \\
\hline
$\bas$ &  $N_0$ & $Q^2_0$ $(\rm{GeV}^2)$ &$m$ $(\rm{GeV})$ & $\ \ \mathrm{\chi^2/d.o.f.}$ \ \ \\ 
\hline
0.1040 $\pm 4.6\times 10^{-4}$ & 0.1311 $\pm 3.7\times 10^{-4}$ & 0.797 $\pm 3.3\times 10^{-3}$ & 0.4743 $\pm 9.6\times 10^{-4}$ & 205.70/166 = 1.239 \\
\hline
0.1100 $\pm 1.8\times 10^{-4}$ & 0.1510 $\pm 8.1\times 10^{-4}$ & 0.809 $\pm 6.2\times 10^{-3}$ & 0.5412 $\pm 1.8\times 10^{-4}$ & 204.73/166 = 1.233 \\
\hline
\hline
\end{tabular}}
\caption{Parameters of the CGC/saturation dipole model: $\bas$, $N_0$,  $Q^2_0$ and  $m$ determined  through fits to the reduced cross-section $\sigma_r$ using the combined H1 and ZEUS data~\cite{HERA1} within the range $0.85 \, \rm{GeV}^2<Q^2<30  \, \rm{GeV^2}$ and $x\leq10^{-2}$. Results are presented for fixed light quark masses $10^{-2}\, \div \,10^{-4}$ and two fixed values of the charm quark masses, 1.40 \rm{GeV} and 1.27 \rm{GeV} respectively}
\label{t1}
\end{table}

Using the parameters of the CGC/Saturation model extracted from the $\chi$-squared fit to the reduced inclusive DIS cross-section, we computed the structure functions $F_2(x, Q^2)$, $F_2^{c\bar{c}}$ for inclusive processes and $\sigma$ vs $Q^2$, $\sigma$ vs $W$, and $B_D$ for exclusive processes, and then compared them to the combined HERA data. The results show that with only four parameters fixed by the reduced cross-section, this model provides a good description of nearly all available data on inclusive and exclusive diffractive processes at HERA for small-x ($x \leq 10^{-2}$). For a comparison of our results with other observables at HERA, refer to Ref.~\cite{SGG}.

 \begin{figure}
 	\begin{center}
	\begin{tabular}{ c c}
 	\leavevmode
 		\includegraphics[width=5.6cm]{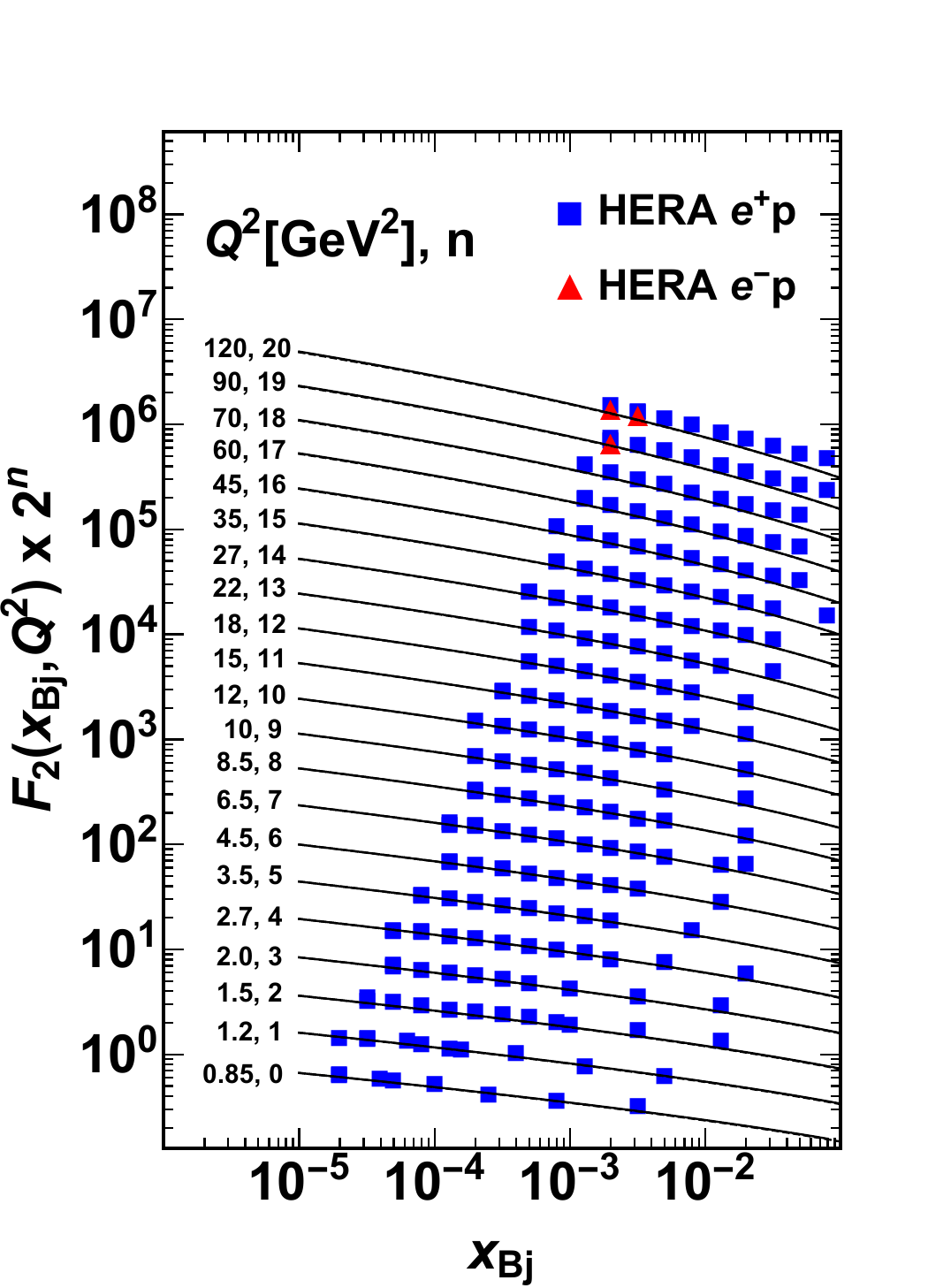}&\includegraphics[width=5.15cm]{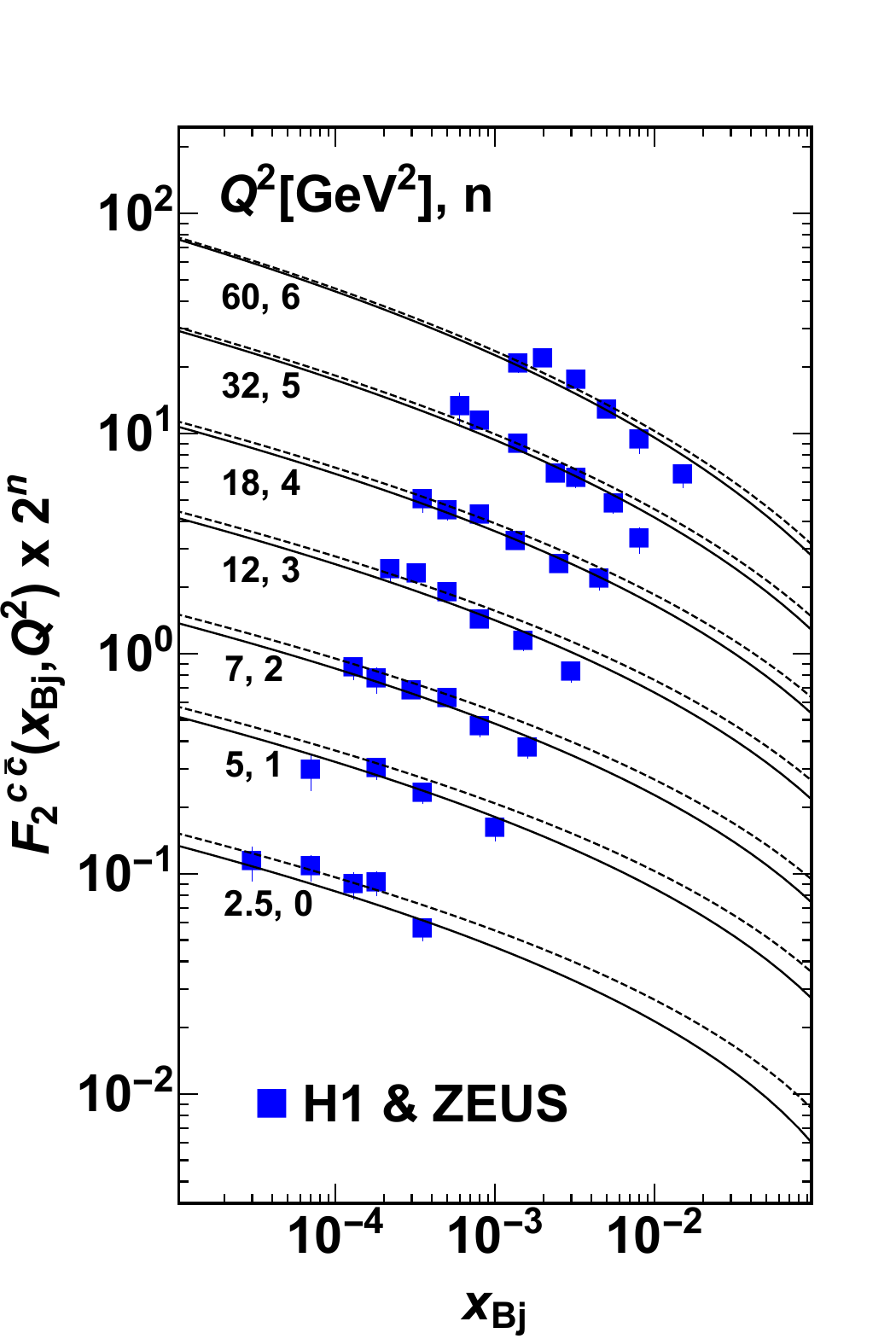} \\
		\fig{nlxs1}-a &\fig{nlxs1}-b\\
		\end{tabular}
			\end{center}
 	\caption{\fig{nlxs1}-a: Structure function $F_{2}(x,Q^{2})$ vs $x$, for different $Q^{2}$ values. Theoretical results and experimental data from H1 and ZEUS~\cite{HERA1}. 		    \fig{nlxs1}-b: Results for charm structure function $F_{2}^{c\bar{c}}(x,Q^{2})$ as a function of $x$, for different values of $Q^2$. The solid lines correspond to parameters from table I with $m_c=1.4\,\rm{GeV}$, and the dashed lines correspond to parameters with  $m_c=1.27\,\rm{GeV}$. The theoretical estimates as well as the experimental data are multiplied by a factor $2^n$ and the values of $n$ are specified in the plot. The experimental data are from H1 and ZEUS collaboration~\cite{HERA2}, assuming $\sigma_r^{c\bar{c}}\,\approx\,F_2^{c\bar{c}}$.}  
 	\label{nlxs1}
 \end{figure}
 
\begin{figure}
\centering 
  \includegraphics[width=15cm]{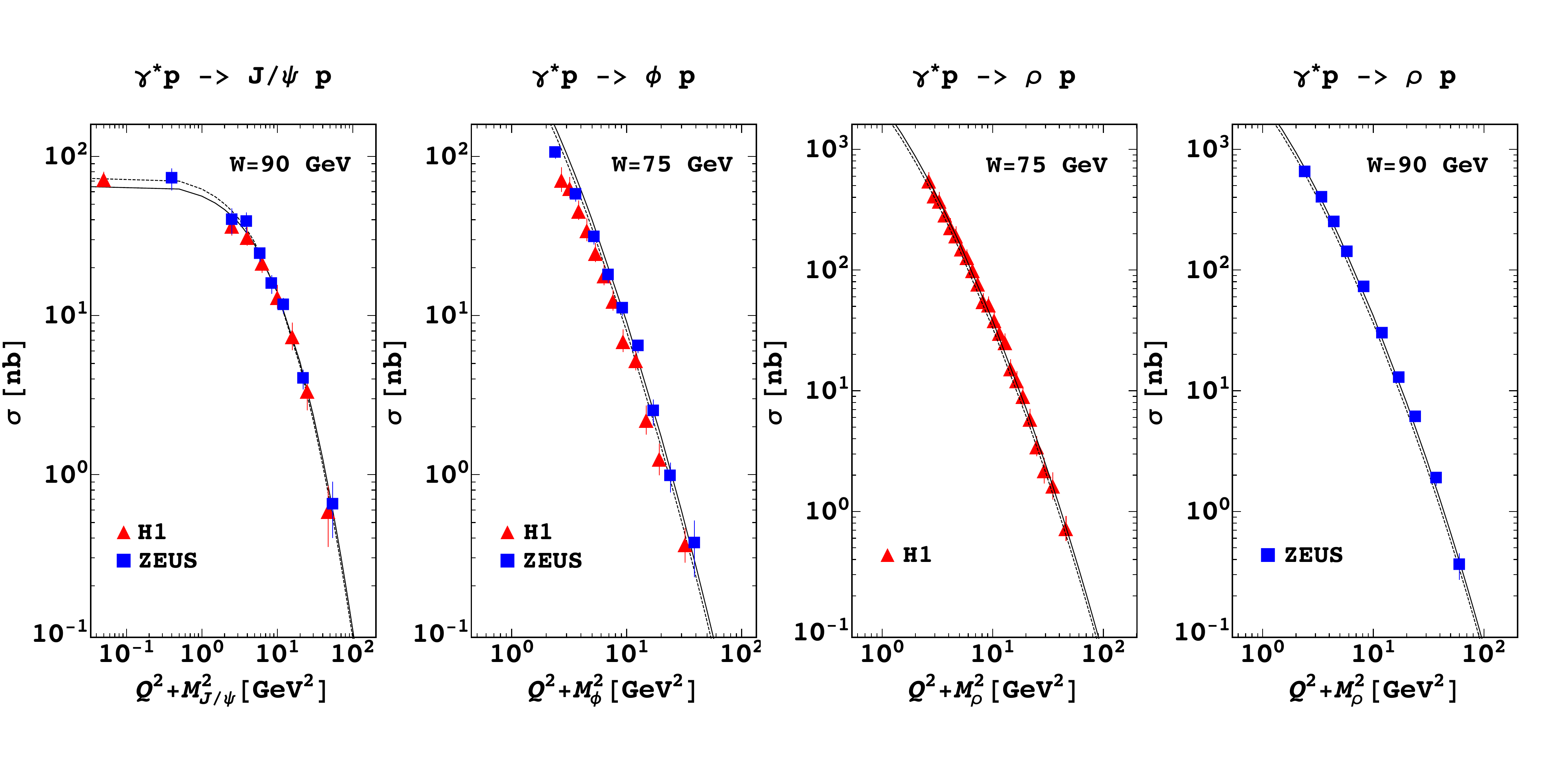}    
\caption{Total vector meson cross-sections $\sigma$ for $J/\psi$, $\phi$ and $\rho$, as a function of $Q^2 + M_{E}^2$ compared to theoretical estimates from CGC/saturation dipole model where solid ($m_c=1.4\,\rm{GeV}$) and dashed ($m_c=1.27\,\rm{GeV}$) lines correspond to the parameters used from table I, respectively. The data are from H1 and ZEUS collaborations~\cite{EXCL1,EXCL2,EXCL3,EXCL4,EXCL5,EXCL6}.}
\label{DVMP-SIGMA-Q2}
\end{figure}

\begin{figure}
\centering 
   \includegraphics[width=14cm]{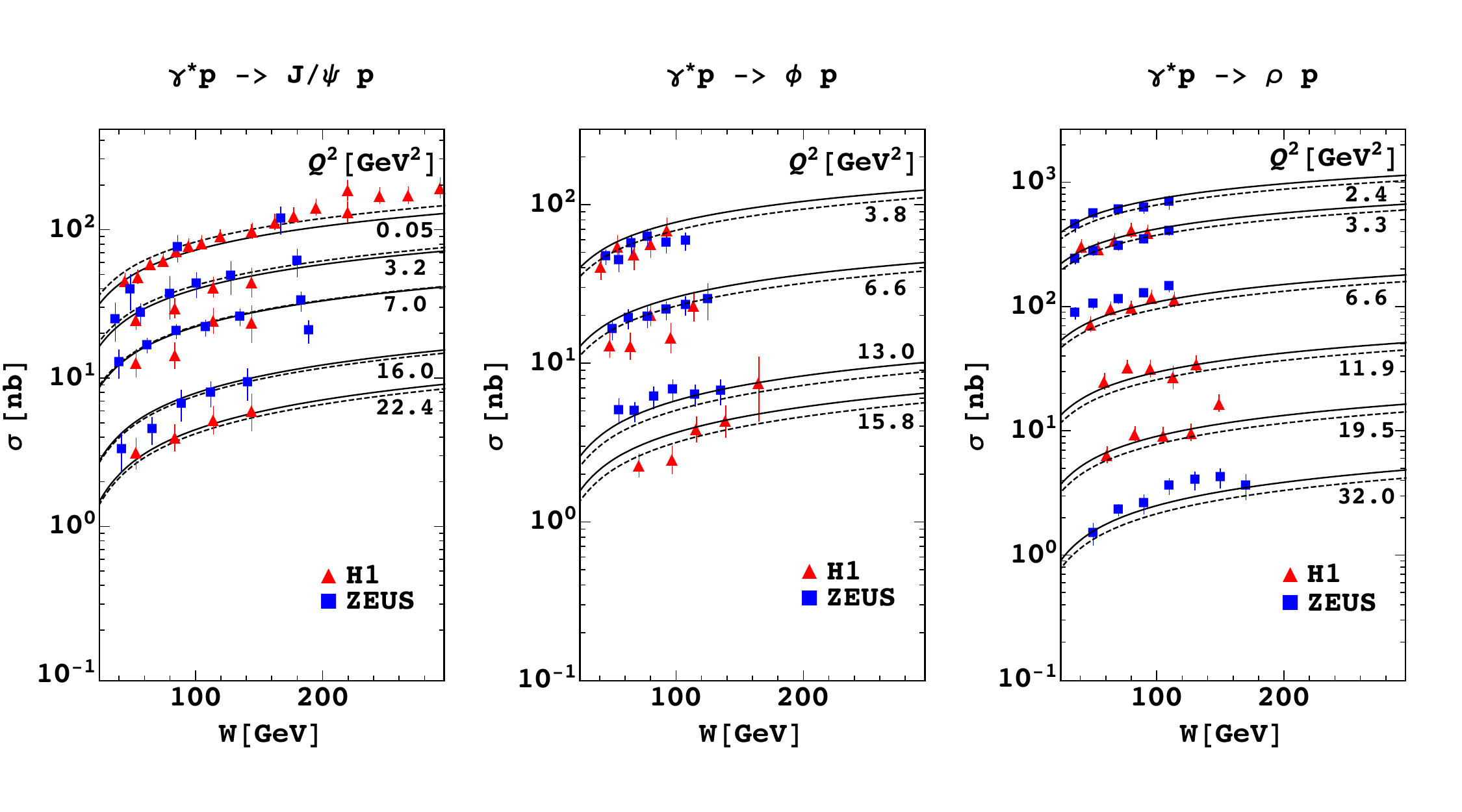}    
\caption{Total vector meson cross-sections $\sigma$ for J/$\psi$, $\phi$ and $\rho$, as a function of $W$. The solid lines
correspond to parameters from table I with  $m_c=1.4\,\rm{GeV}$, and the dashed lines with $m_c=1.27\,\rm{GeV}$. The data are from H1 and ZEUS collaborations~\cite{EXCL1,EXCL2,EXCL3,EXCL4,EXCL5,EXCL6}.}
\label{DVMP-SIGMA-W}
\end{figure}

\begin{figure}
\centering 
   \includegraphics[width=14cm]{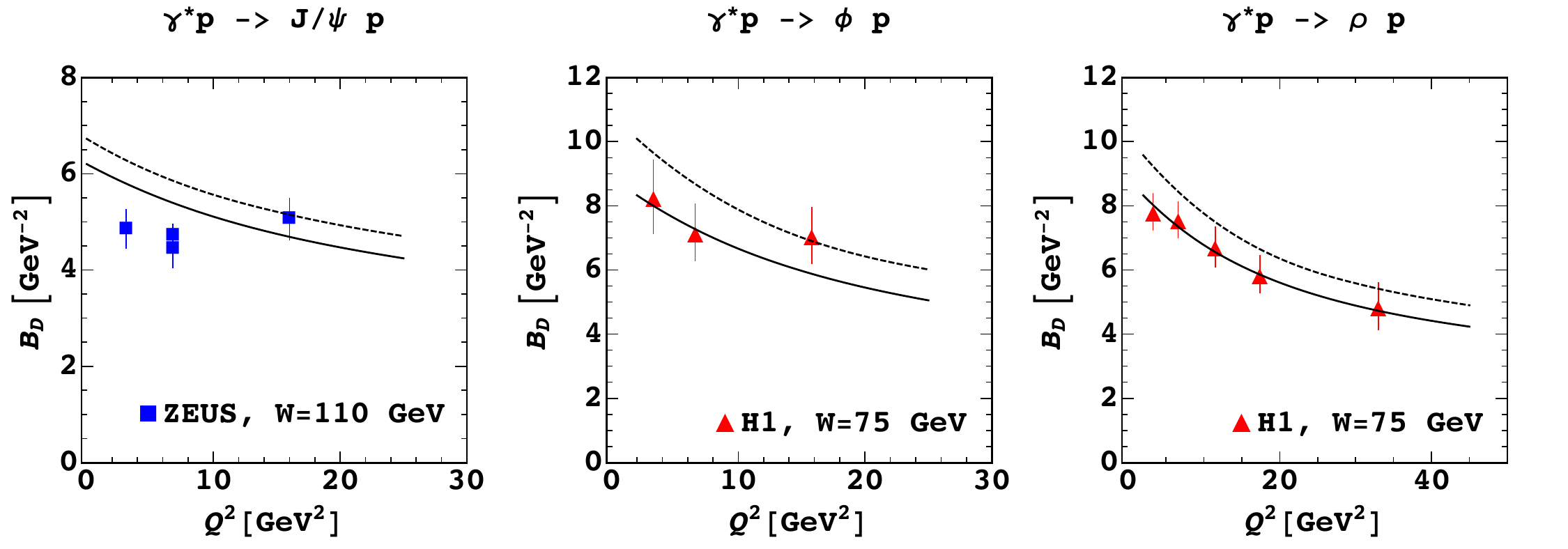}    
\caption{Results for the slope $B_D$ of $t-$distribution of exclusive vector meson electroproduction as a function of $Q^2$. Solid and dashed lines represent calculations using charm quark masses of $m_c = 1.4$ GeV and $m_c = 1.27$ GeV, respectively, from table I. The collection of experimental data are from H1 and ZEUS collaborations~\cite{EXCL1,EXCL2,EXCL3,EXCL4,EXCL5,EXCL6}.}
\label{BDQ}
\end{figure}

%%%%%%%%%%%%%%%%%%%%%%%%%%%%%%%%%%%%%%%%%%%%%%%%%%%%%%%%%%%%%%%%%%%
\section{Summary}
%%%%%%%%%%%%%%%%%%%%%%%%%%%%%%%%%%%%%%%%%%%%%%%%%%%%%%%%%%%%%%%%%%%
To conclude, this study demonstrates the effectiveness of the CGC/saturation dipole model in describing small-x DIS data, incorporating an analytical solution of the BK equation and an impact parameter dependence that satisfies the Froissart theorem. The model achieves good agreement with experimental data across various observables and demonstrates predictive power in describing both inclusive and exclusive processes. This model provides a strong framework for studying high-energy QCD phenomena and guiding future DIS experiments, including those at the upcoming Electron-Ion Collider (EIC).%%%%%%%%%%%%%%%%%%%%%%%%%%%%%%%%%%%%%%%%%%%%%%%%%%%%%%%%%%%%%%%%%%%
%%%%%%%%%%%%%%%%%%%%%%%%%%%%%%%%%%%%%%%%%%%%%%%%%%%%%%%%%%%%%%%%%%%
\section{Acknowledgements}
%%%%%%%%%%%%%%%%%%%%%%%%%%%%%%%%%%%%%%%%%%%%%%%%%%%%%%%%%%%%%%%%%%%
This work was supported by Grant PIIC No. 008/2023 and by PhD scholarship No. 029/2023, from DPP, USM (JG), and by Plan de Fortalecimiento de Universidades Estatales, UPA 19101, CR 18.180, C\'odigo 2390, Universidad de Playa Ancha and ANID Grant No 3230699 (MR). J.G. also thanks to Fondecyt (Chile) grant 1231829 and InES I+D Institutional project, code INID210013/INID2023\_03 of U. of Playa Ancha for financial support to attend the DIS2024 workshop.

\end{document}